\documentclass[runningheads]{llncs}
\usepackage{graphicx}
\usepackage{bbm}
\usepackage{listings}
\usepackage{color}
\usepackage{enumitem}
\usepackage{array}
\usepackage{amsbsy}
\usepackage{amsmath}
\usepackage{xurl}
\usepackage{hyperref}

\definecolor{mygreen}{rgb}{0,0.6,0}
\definecolor{mygray}{rgb}{0.5,0.5,0.5}
\definecolor{mymauve}{rgb}{0.58,0,0.82}
\newcolumntype{P}[1]{>{\centering\arraybackslash}p{#1}}

\begin{document}
\title{Can you See me? On the Visibility of NOPs against Android Malware Detectors} 
%
%
\author{Diego Soi\inst{1}\orcidID{0009-0009-0092-9067} \and
Davide Maiorca\inst{1}\orcidID{0000-0003-2640-4663}
\and
Giorgio Giacinto\inst{1}\orcidID{0000-0002-5759-3017}
\and
Harel Berger\inst{2}\orcidID{0000-0001-6035-5127}}
\authorrunning{Soi et al.}
%
\institute{University of Cagliari, Cagliari (CA) 09123, Italy
\email{\{diego.soi,davide.maiorca,giacinto\}@unica.it} \and
Georgetown University, Washington, DC 20057
\email{bergerar0@gmail.com}}

\maketitle              
\begin{abstract}
Android malware still represents the most significant threat to mobile systems. While Machine Learning systems are increasingly used to identify these threats, past studies have revealed that attackers can bypass these detection mechanisms by making subtle changes to Android applications, such as adding specific API calls. These modifications are often referred to as No OPerations (NOP), which ideally should not alter the semantics of the program. However, many NOPs can be spotted and eliminated by refining the app analysis process. This paper proposes a visibility metric that assesses the difficulty in spotting NOPs and similar non-operational codes. We tested our metric on a state-of-the-art, opcode-based deep learning system for Android malware detection. We implemented attacks on the feature and problem spaces and calculated their visibility according to our metric. The attained results show an intriguing trade-off between evasion efficacy and detectability: our metric can be valuable to ensure the real effectiveness of an adversarial attack, also serving as a useful aid to develop better defenses.

\end{abstract}
\section{Introduction}
\label{sec:introduction}

With its 70\% global market share, Android has become a major target for malware aimed at compromising the devices' confidentiality, integrity, and availability. During the second quarter of 2023, there was a notable increase in malicious software variants, with over 350,000 new instances identified. This surge particularly involved a rise in Ransomware and banking Trojans~\cite{kaspersky23}. 

Extensive research on Android security has led to the development of various Machine Learning-based defences. These systems employ traditional and deep learning techniques, achieving high accuracy and low false positive rates. However, the mutable nature of ML-based systems has inspired adversaries to develop powerful methods to bypass these defences, a concept widely known as Adversarial Machine Learning~\cite{biggio13_ecml,goodfellow15_iclr,Biggio2018_PR}. Research conducted over the last decade demonstrated that a set of subtle, fine-grained alternations in Android executable code or permissions can significantly impair the accuracy of these classifiers. Fortifying learning-based detection systems remains an open and challenging problem.

A critical aspect of these attacks is how they are translated into the \emph{problem space}, i.e., the process of embedding into real samples evasive modifications carried out on the feature level. These changes, often called 'no operations' (or NOPs), are designed not to affect the application's semantics during its execution. NOPs can be implemented in multiple ways: previous works~\cite{pierazzi20_sp} injected portions of benign samples to produce real evasive applications, while others employed redundant or unsatisfied conditions (e.g., fake conditional jumps). Notably, some NOPs possess attributes that make them more perceptible than others to human analysts and machines, thus simplifying the development of countermeasures to detect them (e.g., a sequence of repeated calls to functions that do not perform any action is easily recognizable). Yet, to the best of our knowledge, the impact of the \emph{visibility} and \emph{detectability} of such NOPs remains unexplored. 

Our paper introduces an innovative visibility metric named Clarity Complexity Connection (CCC), aimed at quantifying the ease with which humans can detect NOPs. The CCC metric considers three cardinal components: i) the \emph{predictability} of the injected code line, ii) the \emph{complexity} of the NOP, according to the quantity and diversity of injected instructions, and iii) the NOP's \emph{distinguishability} relative to the application's original codebase.

We applied our metric to various NOPs employed to evade a popular state-of-the-art learning-based system. In particular, we created NOP-based attacks in the problem space to evaluate their visibility. Our results show that effective adversarial attacks should be crafted by considering their stealthiness when developing real evasive applications. In particular, most organizations rely on two layers of security - human experts and detection systems, and the proposed metric can be flexibly employed as a way to consider detection attempts that come both.
Our findings also reveal that the proposed CCC metric can serve as a valuable tool in enhancing and assessing the robustness of malicious detectors against various degrees of NOP injections.

The rest of the paper is organized as follows. First, we describe related work (Section~\ref{sec:SoTa}). Next, we give background on Android and Smali language (Section~\ref{sec:background}). Then, we present our methodology in Section~\ref{sec:methodology}. Our experiments and results are described in Section~\ref{sec:experiments}. Finally, we discuss our insights and future directions (Section~\ref{sec:conclusions}).

\section{Related Work}\label{sec:SoTa}
In this section, we review important studies concerning anti-malware detection techniques such as NOPs (or similar approaches).

NOP or No-Operations involves the insertion of non-functional code or instructions within malicious Android applications. This term mainly concerns explicit NOP commands~\cite{reilly2003no}
that are added to the original code. Christodorescu et al.~\cite{christodorescu2003static} showed that explicit NOPs could evade signature-based malware detection systems. In ADAM~\cite{zheng2012adam} and DroidChameleon~\cite{rastogi2013droidchameleon}, the authors listed explicit NOPs as one of their attack vectors. In addition, Rastogi et al.~\cite{rastogi2013catch} and Faruki et al.~\cite{faruki2014evaluation} implemented explicit NOPs that were effective against several commercial AVs. 

Another attack vector similar to NOPs is the insertion of dead code. Dead code~\cite{knoop1994partial} is program code that can not be executed and is usually obsolete, redundant, or accidentally left in the program, no longer serving any purpose. Aonzo et al.~\cite{Aonzo2020_SoftX} created a tool called ObfuscAPK, in which  (among other attack vectors) dead code was introduced. This instance of dead code comprises mathematical computations and a conditional branch instruction, intentionally structured so that the branch is guaranteed to never be triggered. Lin et al.~\cite{lin2011hunting} suggested inserting codes after a jump API call that will surely be executed. The code after the jump is dead code, as it would not be executed. 

Another similar attack vector is Defunct code/Junk code~\cite{altenberg1994emergent}. Junk code indeed runs (in contrast to the dead code), but does not contribute to the code's functionality. An example can be found in Shankarapani et al.~\cite{shankarapani2011malware}. The authors suggested assembly junk code, in which redundant register mathematical were added to the original code. A similar junk code is also described in the work of Rastogi et al.~\cite{rastogi2013droidchameleon}. Hammad et al.~\cite{hammad2018large} suggested comments, and debugging information as another type of junk code. Bacci et al.~\cite{bacci2018impact} presented an attack of additional registers and the use of them in non-relevant commands. Rosenberg et al.~\cite{rosenberg2018generic} implemented system calls with no effect. Chen et al.~\cite{chen2019android} proposed Android-HIV, in which apps with generated calls to blank functions were introduced. Vasan et al.~\cite{vasan2020mthael} added re-assigning the same value of a variable ($x=x$).


These previous works showed that the NOP addition technique (and similar techniques) can circumvent known malware detectors and AVs. However, as far as we know, none of the previous works has thoroughly explored the visibility of such attacks by considering humans and machines. In other words, these works did not quantify how easy or hard it is to detect and eliminate the NOPs and defunct code.

\section{Android applications}\label{sec:background}




Android applications are packed as zip archives using the APK format~\cite{Kaliciński23_Medium} which contains all resources required by Android to install and execute the application. These consist in several elements among which the most important are (\emph{i}) the \texttt{AndroidManifext.xml} file that holds various information, among which the necessary \emph{permissions} to use restricted functionalities (i.e. API calls), \emph{services}, \emph{actions}, and \emph{components} to run background tasks and communicate with the resource provider; (\emph{ii}) XML files and raw resources (i.e. images, icons, etc.); (\emph{iii}) \texttt{lib} directory to save CPU-specific libraries; and (\emph{iv}) one or more \texttt{.dex} files.

Android apps are often written in Java/Kotlin and compiled in Dalvik bytecode (.dex file), which results from compiling the bytecode throughout a dex compiler. After that, the Android virtual machine DVM (Dalvik Virtual Machine) or ART (Android RunTime), introduced in later Android versions, can execute the application source code.

The reverse process (i.e., disassembling the APK), done by tools like APKtool\footnote{https://apktool.org/}, can extract all APK resources and disassemble .dex files into Smali code, an assembly-like programming language for low-level programming in Android. Smali code is simple to read since it represents the bytecode in a human-readable format.  
Smali syntax is similar to Jasmine's\footnote{\url{https://jasmin.sourceforge.net/}} syntax. The lines are categorized into one of the following: (\emph{a}) \textbf{comments}, which are the notes embedded into the source code starting with $\#$; (\emph{b}) \textbf{directives}, which define classes, methods, registers starting with $.$; (\emph{c}) \textbf{instructions}, which are the bytecode opcodes consisting of a name and one or more parameters.

Listing~\ref{list:JavaCode} shows a simple Java class whose Smali code representation is depicted in Listing~\ref{list:SmaliCode}. Different opcodes are included, and among them, \texttt{sget-\\object} retrieves the value of a static field of type "Object" to put it in register \texttt{v0}; \texttt{invoke-virtual} invokes \texttt{println} non-static method on \texttt{v0} passing \texttt{p0} parameter; eventually \texttt{return-void} indicates the end of a method without returning a specific value.

\begin{lstlisting}[language=Java, caption={Java code defining a class called Demo.}, label={list:JavaCode}]
package com.apk.demo;
public class Demo {
    public static String printMessage(String str) {
        System.out.println(str);
    }
}
\end{lstlisting} 

\begin{lstlisting}[caption={Smali code corresponding to the class defined in Listing \ref{list:JavaCode}. The constructor has been omitted for brevity. Blue strings are comments, red lines are directives, and the cyan lines are instruction opcodes.}, label={list:SmaliCode}, basicstyle=\small]
@.class@ public Lcom/apk/Demo;
@.super@ Ljava/lang/Object;
@.source@ "Demo.java"
@@# direct methods@@
@.method@ public static printMessage(Ljava/lang/String;)V;
    @.param@ p0, "str"
    @.prologue@
    ^sget-object^ v0, Ljava/lang/System;->out:Ljava/io/PrintStream;
    ^invoke-virtual^ {v0, p0}, Ljava/io/PrintStream;->println(Ljava/lang/String;)V
    ^return-void^
@.end@ method
\end{lstlisting}   

\section{Methodology}
\label{sec:methodology}
This section describes our new visibility metric through its three major components (Sections~\ref{subsec:visibility}-\ref{subsec:CCC}),  by giving three representative examples.(Section~\ref{demo}). We also describe the target system (Section~\ref{sec:targetSystem}), and we present the attacker model and several practical evasion attacks on Smali code to assess the utility of the CCC metric (Section~\ref{sec:exampleAttack}).
\subsection{Visibility}
\label{subsec:visibility}
Evasion attacks against learning-based systems should typically feature two major characteristics:
\begin{itemize}
    \item They should be efficient on the \emph{feature space}, i.e., producing an evasive feature vector starting from the original one. Various metrics such as recall, F1, and accuracy can be used to assess the efficacy of these attacks.
    \item Such attacks should also preserve the functionality and semantics of the original samples (i.e., they should be transferred to the so-called problem space~\cite{berger2020evasion,pierazzi20_sp}). In other words, attackers should be able to produce a functional application that implements the feature-space modifications.
\end{itemize}

In this paper, we add another fundamental characteristic these attacks should have: the injected modifications should be \emph{hard to detect} or, in other words, \emph{less visibile} to human perception. Organizations tend to use two layers of security - human experts and ML-based detection systems. This metric paves the way for quantifying the difficulty of human experts to correctly identify attacks (i.e., the \emph{visibility} of the attacks).
Intuitively, various no-operations opcodes scattered through the executable can be a clear sign of manipulation by a human expert. Conversely, a single injected Smali instruction may be indistinguishable from the rest of the code both by humans and machines. More formally, given a malicious sample file $S$, a set of functions $s_i=\{s_1,s_2,s_3...s_n\} \subseteq S$, and a set of lines added to $s_i$ termed $l_i=\{l_1,l_2,l_3...l_n\}$ that an attacker added to evade detection, can the following questions be answered? 

\textbf{RQ1:} How clear is it that the set $l_i$ constitutes a code of attack against a malware detection system and has no meaning? Indeed, some operations, such as explicit NOPs, are clearly an artificial addition to $S$ and regularly, these operations will be described as having no effect.

\textbf{RQ2:} How fast can $l_i$ be identified as a code of attack? Indeed, based on the king of operation encountered when following the functions $s_i$, the time to assess the meaning of the line $l_i$, in the case it is a single line or a loop with multiple computations, may be different.

\textbf{RQ3:} What is the level of connection between $l_i$ to the original $s_i$? If $l_i$ uses original variables from $s_i$, it is difficult to separate $l_i$ from $s_i$. In this case, one usually sees $l_i$ as any other lines of code from $s_i$. On the other hand, if $l_i$ does not include any use of original variables of $s_i$, one can disconnect $l_i$ easily with no great influence on $s_i$ and $S$. In this case, $l_i$ can be identified easily as an unrelated piece of code to $s_i$ and $S_i$. 

To address these questions, we propose a metric called CCC, considering multiple elements. CCC is composed of the following elements:
\begin{enumerate}
    \item Clarity
    \item Complexity
    \item Connection
\end{enumerate}

An evasive set of lines $l_i$ that receives a CCC value $\sim1$ is visible and can be identified as a suspicious tool to evade malware detection. Conversely, a CCC value $\sim0$ is a sign of seemingly invisible evasive code.
Now, we define each element separately. For each element in CCC, we equally split the interval of [0,1] based on representative cases. 
\subsection{Clarity}
\label{subsec:clarity}
Clarity addresses \textbf{RQ1}. This element describes if $l_i$ is clear to be an injected code in $S$. A human expert easily identifies a simple instruction such as explicit NOP as an injected instruction, as it holds no meaning. Conversely, a more complex instruction using variables may seem less clear. Variables conceal the nature of the injected code because they imitate a more meaningful view by a human expert. Two main cases are considered:
\begin{enumerate}
\item If $l_i$ includes an explicit NOP, its clarity rate is considered high, as it is clearly an addition to the original code $S$. As far as we know, no NOP instruction is natively present in Android apps. This is understandable, as NOPs serve no purpose besides evading detection. For this case, $C_1(S)=1$. 
\item If $l_i$ does not include any explicit NOP, it is less clear how to define the clarity element. Thus, we evaluate the average ratio of $l_i$ and the original code lines $s_i$. We note that for small $s_i$, the addition may be more identifiable, as the basic functionality is simple. Therefore, the addition of lines is recognizable. In complex and long functions, the same amount of lines will be less visible. Instead of using the actual amount of added lines $|l_i|$, we use exponential transformation $e^{|l_i|}$, as representing the added lines. Human perception of changes in a codebase can be non-linear. Applying the exponential transformation exclusively to the added lines reflects this non-linear perception, where the perceived impact increases more rapidly with increasing the size of the additions. This aligns with how humans may intuitively assess the significance of the added lines. Also, in larger codebases, a simple linear scaling might lead to saturation, where the modification ratio reaches a ceiling quickly. The exponential transformation helps avoid such saturation, allowing the metric to continue increasing at a more noticeable rate for larger additions. Formally, we compute, 

\begin{equation*}
C_1(S) = \sum_{i=1}^{n}\frac{\frac{e^{|l_i|}}{e^{|l_i|}+|s_i|}}{n}
\end{equation*}
\end{enumerate}

\subsection{Complexity}
\label{subsec:complexity}

Complexity addresses \textbf{RQ2}. This element describes how resourceful an identification of $l_i$ should be. When $l_i$ composes only one-liner perturbation, such as single NOP lines, the complexity is low. A more complex case is with a single function or condition. A condition splits the flow of the code into two options/parts. Therefore, the code analysis now includes two branches, which are more complex to analyze. An even more complex case is with a loop or a nested condition with degree n, as now the analysis may need to do n iterations on $l_i$. The extreme case is reserved for complex computations, such as recursion. The cases are defined as follows:

\begin{enumerate}
\item If $l_i$ is a series of explicit NOPs or single API calls, its complexity rate is considered low. This is because the process needed to identify it is just going over the file and determining the nature of each single line with a small context (before and after the line). For this case, $C_2(l_i)=0$. 
\item If $l_i$ is a 
single function or a conditional, the process of determining its nature is more complex. An analysis of the function operation or the branches of the condition is needed. In this case, a human expert needs to have a simulated memory of variables' values and the connections between variables to be able to determine the nature of $l_i$. Therefore, in this case, $C_2(l_i)=0.33$.

\item The next value is assigned for $l_i$ when it is a loop or a nested condition with degree n.
This case sometimes requires iterating on the code a couple of times, thus resulting in a higher complexity rate. For this case, $C_2(l_i)=0.66$.
\item The last and extreme value is assigned for complex computational tasks, such as recursion or a series of nested loops. Following recursion, for example, is a tough assignment for human experts, and therefore is assigned with the maximum values. For this case, $C_2(l_i)=1$. 
\end{enumerate}

We define $C_2$ to be the complexity element of a test sample file $S$ as follows: 

\begin{equation*}
C_2(S) =\frac{\sum_{i=1}^{n}C_2(l_i)}{n}
\end{equation*}

\subsection{Connection}
\label{subsec:comprehension}

Connection addresses \textbf{RQ3}. This element addresses the difficulty of detachment of the code snippet $l_i$ from $s_i$. A simple set $l_i$ that is not connected to parts of the original functions $s_i$ is easy to remove. Consequently, in this case, $l_i$ can be viewed as an external part of $s_i$. On the other hand, the set $l_i$ that uses original variables is more difficult to remove, as it needs a more in-depth connection of the $s_i$. Three cases are considered:

\begin{enumerate}
\item The connection value is low if there are no real attachments from $l_i$ to $s_i$, such as in the case of explicit NOP or API calls without variables. Another case is if the set $l_i$ constitutes a function that is not called from the original code. Thus, in these cases, $C_3(l_i)=0$.
\item If $l_i$ includes one original variable from $s_i$, the $C_3$ value is higher, as now $l_i$ "seems" interconnected to $s_i$. A human expert needs to follow the modifications to this variable during $l_i$ to be able to separate the set $l_i$ from the original code $s_i$. Thus, in this case, $C_3(l_i)=0.5$.
\item For the last case, $l_i$ incorporates more than one original variable from $s_i$. Thus, it may seem as if $l_i$ is indistinguishable from the original code of $s_i$. This case is different from the last case, as in this current case, a human expert needs to keep in mind the values of different variables and watch their changes during the simulation of the code. Following only one variable is easier. Therefore, in this case, we determine that $C_3(l_i)=1$.
\end{enumerate}

We define $C_3$ to be the connection element of a sample file $S$ as follows: 

\begin{equation*}
C_3(S) =\frac{\sum_{i=1}^{n}C_3(l_i)}{n}
\end{equation*}

\subsection{CCC}
\label{subsec:CCC}
The notion of visibility is to quantify how hard it is to identify an attack. Therefore, a high complexity ($C_2$) value contributes to a more invisible attack, suggesting more resources to acknowledge the attack. Also, a high value of connection ($C_3$) determines that it is more difficult to distinguish between the original code lines and the new code lines. Therefore, the CCC formula uses the values of $1-C_2(S)$ and $1-C_3(S)$ instead of the positive values. 

Additionaly, a set of weights $\{w_i:i=1..3\}$ is assigned for the three Cs. A naive set of weights is \{0.33, 0.33, 0.33\}. However, it is not a clear cut to give each part the same weight. For example, a long list of consecutive NOPs, also known as NOP sled~\cite{akritidis2005stride}, can be easily identified. This is true, as even if a software developer mistakenly added a single NOP to its code, the probability that he will do so with a long list of NOPs is negligible. Using a long list of NOPs is too suspicious from being a natural part of an app. It can be a part of an original malicious app or an addition to a benign one. In both cases, the list of NOPs tends to signal malicious activity. Therefore, $C_1$ plays a vital part in addressing visibility and should be assigned with a reasonable weight.

Also, if a set of code lines does not have any connection to the code (i.e., $1-C3(S)\sim1$, is is reasonable to say that the code is redundant and does not run at all (a.k.a. junk code). A simple example is a non-called function added to the original code. This is a red light for manipulating the code, since a product version of an app regularly does not include redundant functions. As a consequence, $C_3$ should get a reasonable weight. 

However, the complexity value $C_2$ aspect does not have clear cuts as the other two aspects. This aspect relates to the complexity of the code snippet. A very complex code or a very subtle code does not necessarily state anything about its nature. It is just the way the code is organized. However, as a human expert's ability to identify a malicious code snippet is affected by its complexity, we included it in the CCC metric.

Consequently, the $w_1$ and $w_3$ receive a higher value than $w_2$. Now, we define the CCC metric:
\begin{equation*}
CCC(S) = w_1*C_1(S)+w_2*(1-C_2(S))+w_3*(1-C_3(S))
\end{equation*}

Given the above-mentioned observations on the differences between the weights, we pick the values of 0.4,0.2,0.4 for $w_i$, respectively. Other weights can be picked as well. This analysis is left for future work. Thus, our final version of the CCC visibility metric is:

\begin{equation*}
CCC(S) = 0.4*C_1(S)+0.2*(1-C_2(S))+0.4*(1-C_3(S))
\end{equation*}

\subsection{Representative Examples}
\label{demo}
We present three representative code snippets showing different attack scenarios to demonstrate our innovative CCC metric. The demo evaluation is summarized in Table~\ref{tab:tab_d}. We start with two simple functions on two integers - addition and subtraction (Listings~\ref{list:OriginalCode}):

\begin{lstlisting}[caption={Small code snippet with addition and subtraction functions. Red lines are directives}, label={list:OriginalCode}, basicstyle=\small]
@.method@ public static addTwoIntegers(II)I;
    @.registers@ 3
    add-int v0, v1, v2
    return v0
@.end @method

@.method@ public static subtractTwoIntegers(II)I;
    @.registers@ 3
    sub-int v0, v1, v2
    return v0
@.end @method
\end{lstlisting}

In the following, we report three examples that are based on these simple functions.
\subsubsection{Simple NOPs (EX\_1)}
\label{s_nops}
In this example (Listings~\ref{list:NopAddition}), we added a set of three NOPs to two original functions. The $C_1$, $C_2$, $C_3$ values for $EX_1$ are 1, 0, 0, respectively, and the final CCC value is 1. The full calculation is available in Appendix~\ref{appendixA-simple}. The high CCC value pinpoints that $EX_1$ is pretty simple to distinguish from the original code. This is understandable, as the code snippet includes only NOPs, a suspicious set of lines.

\vspace{1em}
\begin{lstlisting}[caption={Small code snippet with simple NOPs additions. Red lines are directives, cyan lines are additions to the code.}, label={list:NopAddition}, basicstyle=\small]
@.method@ public static addTwoIntegers(II)I;
    @.registers@ 3
    add-int v0, v1, v2
    ^nop^
    ^nop^
    ^nop^
    return v0
@.end @method

@.method@ public static subtractTwoIntegers(II)I;
    @.registers@ 3
    sub-int v0, v1, v2
    ^nop^
    ^nop^
    ^nop^
    return v0
@.end @method

\end{lstlisting}

\subsubsection{Loop \& Addition (EX\_2)}
\label{loop_nops}
In this example (Listings~\ref{list:loopNop}), we used the same simple functions as before. We added a loop to the original code with a simple addition. The $C_1$, $C_2$, $C_3$ values for $EX_2$ are 0.98, 0.66, 1, respectively, and the final CCC value is 0.46. The full calculation is available in Appendix~\ref{Appednix-loop}. The CCC value of 0.46 implies that $EX_2$ demonstrates a slightly tricky code, that is not trivial to distinguish from the original one. The small amount of original code contributes to the magnitude of $C_1$, therefore the value is not negligible. For longer functions, the addition of only 5 lines has a smaller effect. For simplicity, we demonstrated the metric on a small code snippet. This code snippet includes an advanced connection to the original code. An in-depth understanding of the loop is needed to analyze it correctly. 
\vspace{1em}

\begin{lstlisting}[caption={A code snippet with simple loop and mathematical addition. Red lines are directives, cyan lines are additions to the code.}, label={list:loopNop}, basicstyle=\small]
@.method@ public static addTwoIntegers(II)I;
    @.registers@ 3
    add-int v0, v1, v2
    ^const/4 v1, 0^  @@# Initialize loop variable to 0@@
    ^:start_loop^
    @@# Check if the loop variable is less than v0@@
    ^if-ge v1, v0, :end_loop^
    @@# Add v2 to itself and store the result in v2@@
    ^add-int v2, v2, v2^
    @@# Increment the loop variable@@
    ^add-int/lit8 v1, v1, 1
    goto :start_loop^
    ^:end_loop^
    return v0
@.end @method

@.method@ public static subtractTwoIntegers(II)I;
    @.registers@ 3
    sub-int v0, v1, v2
    ^const/4 v1, 0^  @@# Initialize loop variable to 0@@
    ^:start_loop^
    @@# Check if the loop variable is less than v0@@
    ^if-ge v1, v0, :end_loop^
    @@# Add v2 to itself and store the result in v2@@
    ^add-int v2, v2, v2^
    @@# Increment the loop variable@@
    ^add-int/lit8 v1, v1, 1
    goto :start_loop^
    ^:end_loop^
    return v0
@.end @method
\end{lstlisting}  

\vspace{1em}
\begin{lstlisting}[caption={A code snippet with simple condition and API call (print). Red lines are directives, cyan lines are additions to the code.}, label={list:NopCondition}, basicstyle=\small]
@.method@ public static addTwoIntegers(II)I;
    @.registers@ 3
    add-int v0, v1, v2
    @@# Compare v2 with 0@@
    ^if-ge v2, 0, :skip_print^
    @@# Code to print nothing (invoke-static)@@
    ^invoke-static {}, Ljava/lang/System;->outPrintln()V^
    ^:skip_print^
    return v0
@.end @method

@.method@ public static subtractTwoIntegers(II)I;
    @.registers@ 3
    sub-int v0, v1, v2
    @@# Compare v2 with 0@@
    ^if-ge v2, 0, :skip_print^
    @@# Code to print nothing (invoke-static)@@
    ^invoke-static {}, Ljava/lang/System;->outPrintln()V^
    ^:skip_print^
    return v0
@.end @method

\end{lstlisting}   

\subsubsection{Condition \& API Call (EX\_3)}
\label{condition_nops}
In this example (Listings~\ref{list:NopCondition}), we used the same original functions as well. We added to the original code a condition with a simple API call of print (with no content). $C_1$, $C_2$, and $C_3$ values for $EX_3$ are 0.79, 0.33, 0.5, respectively, and the final CCC value is $0.65$. The full calculation is available in Appendix~\ref{Appenidx-condition}. The CCC value of 0.65 implies that $EX_3$ is less invisible than $EX_2$. This is because the condition makes it harder to analyze on the one hand, and the API call makes it easier as it does not include any use of original variables. Thus, only one original variable is used, which makes it less connected to the original code.




\begin{table}[h!]
\centering
\caption{CCC demo examples. The weights for $C_1$, $1-C_2$, \& $1-C_3$ are 0.4, 0.2, 0.4, respectively.}
\begin{tabular}{|P{1cm}|P{1cm}|P{1.5cm}|P{1.5cm}|P{1.5cm}|}
\hline
\rule{0pt}{3ex}      & \textbf{\pmb{$C_1$}}  & \textbf{\pmb{$1-C_2$}} & \textbf{\pmb{$1-C_3$}} & \textbf{\pmb{$CCC$}} \\[2pt] \hline
\rule{0pt}{3ex} \textbf{\pmb{$EX_1$}} & 1    & 1    & 1    & 1     \\[2pt] \hline
\rule{0pt}{3ex} \textbf{\pmb{$EX_2$}} & 0.98    & 0.66  & 1    & 0.46   \\[2pt] \hline
\rule{0pt}{3ex} \textbf{\pmb{$EX_3$}} & 0.79 & 0.33 & 0.5 & 0.65  \\[2pt] \hline
\end{tabular}
\label{tab:tab_d}
\end{table}
\vspace{-1em}
\subsection{Target System}
\label{sec:targetSystem}
The target machine we used was proposed by McLaughlin et al.~\cite{McLaughlin2017_ACM}, a DL detection system based on a Convolutional Neural Network. The system's features are opcode sequences statically extracted by reading the disassembly Smali file. These are represented as one-hot vectors, projected into an embedding space to generate a matrix $P$ with dimensions $n \times k$, where $k$ is the embedding space's dimensionality, and $n$ is the number of original opcodes.

The detection system consists of 1D CNN layers, with the first one taking the $n \times k$ matrix as input. The output of the last CNN layer undergoes a max pooling layer, followed by fully-connected layers. Eventually, a softmax function computes the classification score of benign and malware classes (i.e., classes 0 and 1).

\subsection{Evasion Attacks}\label{sec:exampleAttack}

In this section, we present the attacker model and three evasion attacks we implemented to assess the practicality of the CCC metric. 

\subsubsection{Attacker Model}
\label{sec:at_model}
The attacker in this work has access to the pre-processing that is done to extract the opcodes and feature representation (i.e., opcodes one-hot encoding) of the target system. Thus, the attacker can be considered a gray-box attacker.

\begin{figure}[h]
    \centering
    \makebox[\textwidth][c]{\includegraphics[scale=0.45]{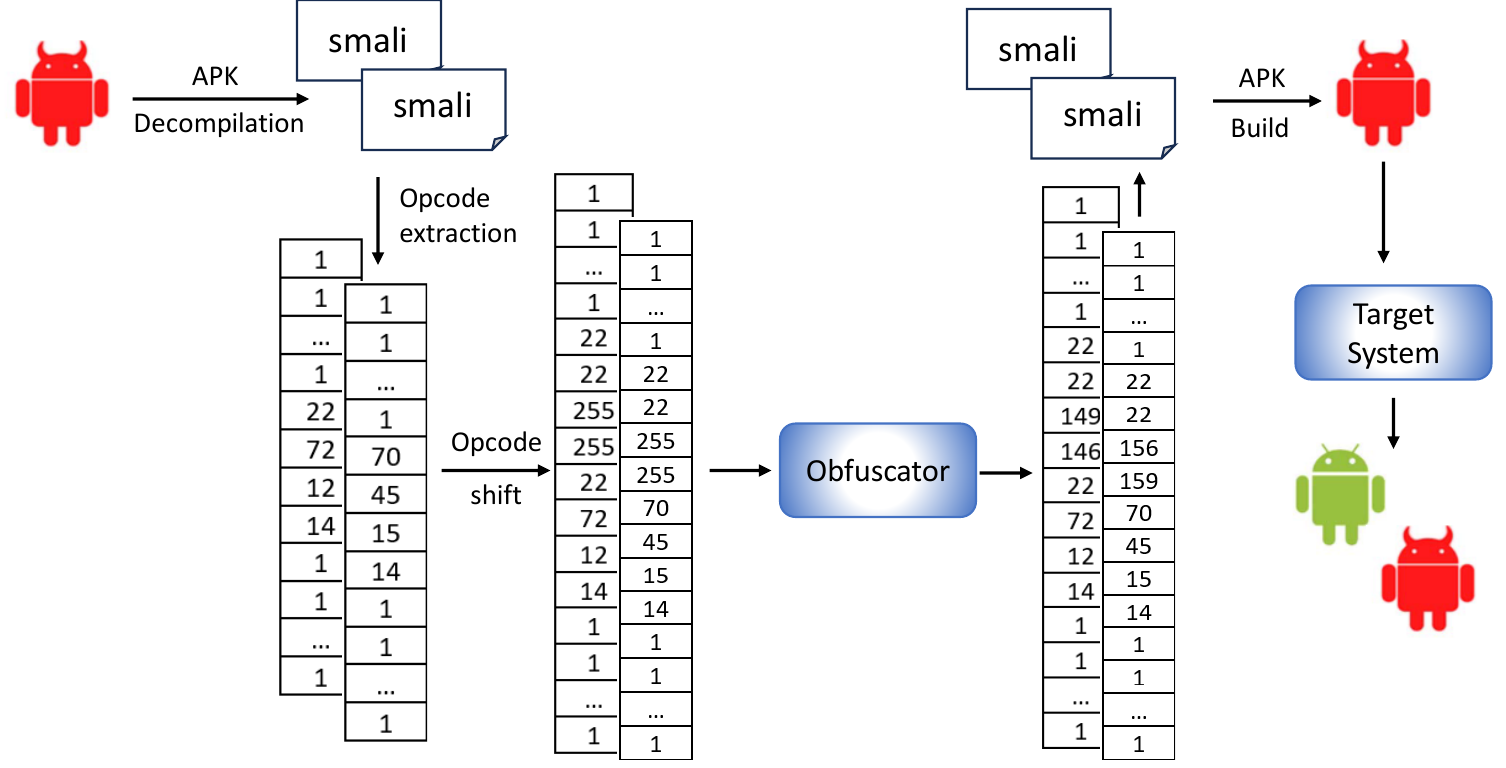}}
    \caption{The attack process for SIO attack. Opcode sequences contain $1$ as padding between Smali functions. After opcode shifting, sequences contain additional opcodes (i.e. 22, 22, 255, 255 representing \texttt{const, const, x, x}). The attack will find the best opcodes reducing the maliciousness score to change the injected "255". Note that IMI attack works similarly but the added opcodes represent \texttt{const, if-eqx, x, x}.}
    \label{fig:attackDescription}
\end{figure}
\vspace{-1em}
\subsubsection{Attacks}
\label{sec:attacks}
The three attacks we implement are based on the instrumentation of Obfuscapk (a Python modular tool used to obfuscate APKs in different ways~\cite{Aonzo2020_SoftX}) and SecML (an automatic framework to create feature-space adversarial examples~\cite{demetrio2021secmlmalware}). In particular, we created a custom injection tool capable of creating fully working APKs modified through the injection of single smali instructions. More specifically,  the attacks are implemented as follows:
\begin{enumerate}
    \item \textbf{Simple NOP:} The first attack is simple and injects regular NOP instructions inside the Smali code. That is performed by employing Obfuscapk on the test samples.
    \item \textbf{Simple Opcode \& Impossible If:} The other two attacks are a shift attack on each Smali code method that is analyzed by the target system (i.e., all Smali files up until $8192$ opcodes). 
The attack process is shown in Figure~\ref{fig:attackDescription}.
The steps for these attacks are as follows: First, we extract the opcodes like the target system, shifting them to add more instructions based on the attack variants. Then, we employed SecML optimizer and a modified version of the discretize byte evasion~\cite{Demetrio2021_ACM,Kolosnjaji2018_EUSIPCO} to select opcodes \texttt{x} so that the confidence of the malware class drops under a certain threshold. 
The two attack variants are:
\begin{enumerate}
    \item \textbf{Simple opcode attack (SIO):} \texttt{const, const, x, x}.
    \item \textbf{Impossible if attack (IMI):} \texttt{const, if-eqz, x, x}
\end{enumerate}
Listings~\ref{list:simpleOpcode} and~\ref{list:impossibleIf} demonstrate the use of the simple opcode and impossible if attacks.
\vspace{1em}

\begin{lstlisting}[caption={A code snippet with simple opcode attack (SIO). Red lines are directives, cyan lines are additions to the code.}, label={list:simpleOpcode}, basicstyle=\small]
@.method@ public static addTwoIntegers(II)I;
    @.registers@ 4      # inserting one more register
    ^const v0, 0x8^
    ^const v3, 0xA^
    ^sub-int v0, v0, v3^
    ^xor-int v0, v0, v3^
    add-int v0, v1, v2
    return v0
@.end @method
\end{lstlisting} 

\begin{lstlisting}[caption={A code snippet with impossible if (IMI). Red lines are directives, cyan lines are additions to the code.}, label={list:impossibleIf}, basicstyle=\small]
@.method@ public static addTwoIntegers(II)I;
    @.registers@ 3
    ^const v0, 0x1^
    ^if-eqz v0, :impossible^
        ^sub-int v0, v1, v2^
        ^xor-int v0, v1, v2^
    ^:impossible^
    add-int v0, v1, v2
    return v0
@.end @method
\end{lstlisting} 
     
\end{enumerate}

\section{Experimental Details}\label{sec:experiments}
In this section, we present the dataset, the two experiments, and the experimental evaluation.

\subsection{Dataset} 
The dataset consists of apps from Androzoo and VirusShare. Specifically, Androzoo samples were labeled using VirusTotal APIs, considering the number of sandboxes and antivirus vendors identifying them as malicious. Specifically, if at least two of them flagged the app as malware, we labeled it accordingly. The dataset encompasses around $10000$ applications (i.e., $5000$ APKs per category) developed from $2008$ to $2022$ whose hashes are listed in \href{https://anonymous.4open.science/r/NOPs_Visibility-146A}{Anonymous GitHub}. We divided the dataset into training and test sets at an 80/20 ratio. 

\subsection{Experiments} 
To measure the effectiveness of the attacks, we ran an experiment in which we manipulated a set of random $\sim1000$ malicious applications by each one of the attacks. All three attacks were constructed employing ObfuscAPK and SecML. We reconstructed all manipulated Smali files to build a working APK. Table~\ref{tab:results} presents the classification and visibility results.
We also run the simple opcode attack with different lengths of injected opcode set. We tested this to see how the visibility value changes according to different amounts of injected content. We tested the same applications for this experiment as well.

Additionally, to assess whether the obfuscated APKs are still functional, we tested $\sim30\%$ of the manipulated apps using an emulated environment employing Android Studio and Droidbot~\cite{Li2017_ICSE}, a test input generator for Android. The testing involves running the applications for one minute to discern any discrepancies between the original applications and their modified counterparts. The results are then analyzed to confirm that the obfuscated APKs are functioning as expected. 
\vspace{-1em}
\subsection{Results - Exp 1: Assessing visibility}\label{sec:visibilityResults}
The goals of this experiment are to assess how specific NOP-based attacks affect the classifier's decision, and to understand what is the visibility value.

\noindent \textbf{Simple NOP attack} In the case of the NOP attack, as described in Section~\ref{s_nops}, the $CCC$ value is equal to $1$ since it is easy, even automatically, to recognize the presence of the injection and possibly eliminate it. Running the attack on our target system, resulted in an accuracy rate of 0.92, precision rate of 0.89, recall rate of 0.93, and f1-score of 0.91. 

\noindent \textbf{SIO attack}. 
For the simple opcode attack (SIO) exampled in Listings~\ref{list:simpleOpcode}, the CCC value is computed as follows:
$C_1(SIO)=0.82$, as the additional lines are only 4 lines in comparison to the number of original method lines (i.e., 12 on average). This computation has been done, averaging the clarity on all the modified samples. 
$C_2(SIO)=0$, because its complexity rate is low. $C_3(SIO) = 1$, as there is more than one original variable in use. Thus, $CCC(SIO)=0.53$.
The SIO attack against the target system got an accuracy rate of 0.51, a precision rate of 0.68, a recall rate of 0.24, and an f1-score of 0.35.

\noindent \textbf{IMI attack}. Listing~\ref{list:impossibleIf} shows an example of the IMI attack we performed. In this case, we can compute the CCC value as follows:
$C_1(IMI)=0.82$, as there are 4 additional lines (excluding the tag) out of the total number of original method lines (i.e., 12 on average). $C_2(IMI)=0.33$, as there is a use of a conditional instruction.
$C_3(IMI) = 1$, as more than one original variable is used. Thus,  $CCC(IMI)=0.46$.

The IMI attack against the target system got an accuracy rate of 0.59, a precision rate of 0.77, a recall rate of 0.38, and an F1-score of 0.51.
Table~\ref{tab:results} summarizes the classification and visibility results. The results highlight a notable decrease in metrics during the performed attack. Specifically, the accuracy rate drops by 32\%-45\%, and the recall rate declines by 48\%-69\%, suggesting a significant reduction in the overall effectiveness of recognizing malicious samples during the attack.

Choosing the third type of attack (i.e., \emph{IMI}) might be more favorable for evading human detectability. Despite having a higher recall than the SIO attack, the CCC value is slightly lower. 
The SIO attack seems more effective in means of recall and, as a consequence, a better competitor against the detection system. However, as the CCC value shows, its visibility value is higher. Therefore, it is less effective in a setting of both a detection system and a human expert than in an IMI attack.

\begin{table}[h]
    \centering
    \caption{Classification and visibility results. The metrics are precision, recall, f1-score, and our CCC visibility metric.}
    \begin{tabular}{|P{2cm}|P{1.5cm}|P{2.75cm}|P{2.75cm}|P{2.75cm}|}
    \hline
    \rule{0pt}{3ex}          & \textbf{Original APKs}   & \textbf{Simple NOP} & \textbf{Simple Opcode} & \textbf{Impossible IF} \\[2pt] \hline
    \rule{0pt}{3ex} \textbf{CCC}    & -       & 1       & 0.53     & 0.46\\[2pt]\hline
    \rule{0pt}{3ex} \textbf{Accuracy} & 0.92    & 0.60    & 0.51    & 0.59     \\[2pt] \hline
    \rule{0pt}{3ex} \textbf{Precision} & 0.89   & 0.85  &  0.68   & 0.77   \\[2pt] \hline
    \rule{0pt}{3ex} \textbf{Recall}   & 0.93 & 0.45 & 0.24 & 0.38  \\[2pt] \hline
    \rule{0pt}{3ex} \textbf{F1-score} & 0.91 & 0.61 & 0.35 & 0.51  \\[2pt] \hline
    \end{tabular}
    \label{tab:results}
\end{table}
\vspace{-1em}
\subsection{Results - Exp 2: CCC vs Recall}\label{sec:lengthsResults}
\begin{figure}[h]
    \centering
    \includegraphics[scale=0.5]{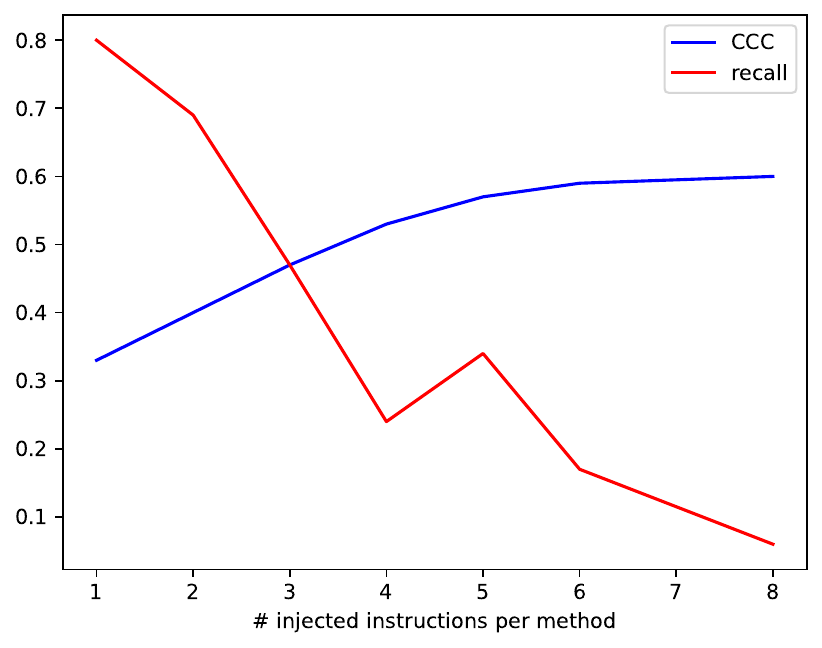}
    \caption{Relationship between CCC value and recall with SIO attack. The values are the mean of the recall and CCC values for the test samples.}
    \label{fig:results}
\end{figure}

A second experiment explored the relationship between CCC value and recall by employing the simple opcode attack with varying numbers of injected opcodes. Figure~\ref{fig:results} illustrates the findings, revealing a kind of proportionality between visibility and the number of injections. Indeed, as explained in Section~\ref{subsec:visibility}, a larger amount of injected opcodes increases the opportunity of a human expert to identify the attacks (i.e. a higher visibility value). Additionally, as expected, the recall decreases because the more opcodes chosen by the attacker, the lower the model's capability to recognize malicious samples is.

One would have expected that the CCC metric would have a minor effect on the recall. This is a reasonable assumption, as the more opcodes injected in the app, the more non-malicious it seems in the view of the detection system and more visible in the view of a human expert. However, as we constructed our metric, the CCC value was not solely computed by the amount of code injected into the sample in the clarity element $C_1$, but also by the complexity of the code $C_2$ and the connection between the injected and the original one $C_3$ which normalize the increase in CCC value. Still, there is an increase in visibility, which confirms our theory that more injected code means less recall and higher CCC.

\section{Discussion and Conclusion}\label{sec:conclusions}


This paper proposes a new visibility metric called CCC, which quantifies the difficulty required to detect evasion adversarial examples. A common misconception in the security community is to focus on strengthening the detection systems against new threats without considering the human aspect. Based on our novel metric, it is clear that both aspects of security should be addressed, as attacks can overcome both. We analyze various evasion attacks with traditional metrics and our CCC value drawing insights considering the tradeoff between the efficacy, in terms of recall, and the visibility of the attack, showing that a more effective attack is the one with a good combination between CCC and recall. Our findings underscore the importance of the CCC metric in bolstering a two-layered defence system, combining ML-based detection with human expertise.

Our metric paves the way for understanding the connection between a human expert's view of a manipulated sample and the view of a detection system on the same subject. However, our research has some limitations. First, we did not fully evaluate the different weights of the three elements of CCC. Optimization of the different components of CCC is an open issue that we aim to explore in future work. 
Additionally, we want to expand the variety of detection systems (from both academia and the industry) and evasion attacks in Android malware detection to get a broader view of the impact of our metric in this domain. 
At last, we aim to extend our metric's applicability to other malware detection domains, like PDF and PEs, which will be explored in future works.

The threat of evasion attacks is yet to be mitigated. More attacks are devised daily. We hope our new metric will aid in correctly mapping the risk levels of attacks and train both human and ML-based detection systems against these attacks. 

%
%
%
\vspace{0.5em}
\noindent\textbf{Acknowledgements}. This work was partially supported by project SERICS (PE00000014) under the NRRP MUR program funded by the EU - NGEU.
\bibliographystyle{splncs04}
\bibliography{bibliography}

\appendix
\section{Full Calculations - Code Examples (Section~\ref{demo})}
\label{appendixA}
In this appendix, we show the full calculations of the examples in Section~\ref{demo}. 
\subsection{Simple Nops}
\label{appendixA-simple}
The following calculations refer to the addition of three NOPs.

For $C_1(EX_1)$, the value is 1, as $EX_1$ includes explicit NOPs (at least one). This is a red flag for manipulation of an original code.

For $C_2(EX_1)$, we get:
\begin{equation*}
C_2(EX_1) =\sum_{i=1}^{n}\frac{C_2(l_i)}{n}=\frac{0}{2}+\frac{0}{2}=0
\end{equation*},
as EX1 includes only NOPs.

For $C_3(EX_1)$, we get:
\begin{equation*}
C_3(EX_1) = \sum_{i=1}^{n}\frac{C_3(l_i)}{n}= \frac{0}{2}+\frac{0}{2}=0
\end{equation*}

Thus, the CCC of EX1 is:
\begin{eqnarray*}
    CCC(EX_1) &= w_1*C_1(EX_1)+w_2*(1-C_2(EX_1))+w_3*(1-C_3(EX_1))
    \\ &= 0.4 * 1 +0.2*(1-0)+0.4*(1-0) = 1 
\end{eqnarray*}

\subsection{Loop \& Addition (EX\_2)}
\label{Appednix-loop}
The following calculations refer to a loop with simple addition. For these calculations, we consider the lines of the constant, loop increment, condition, and additional additions (the tags of the loops are eliminated from the computations, as they are basically semantic pointers). Thus, the additional code in each function is 5 lines, and the original code in each function consists of 2 lines.

For $C_1(EX_2)$, we get:
\begin{equation*}
C_1(EX_2) = \sum_{i=1}^{n}\frac{\frac{e^{|l_i|}}{e^{|l_i|}+|s_i|}}{n} = \frac{\frac{e^5}{e^5+2}+\frac{e^5}{e^5+2}}{2}=0.98
\end{equation*}

For $C_2(EX_2)$, as EX2 includes a loop, we get:
\begin{equation*}
C_2(EX_2) =\sum_{i=1}^{n}\frac{C_2(l_i)}{n}=\frac{0.66}{2}+\frac{0.66}{2}=0.66
\end{equation*}

For $C_3(EX_2)$, we get:
\begin{equation*}
C_3(EX_2) =\sum_{i=1}^{n}\frac{C_3(l_i)}{n}=\frac{1}{2}+\frac{1}{2}=1
\end{equation*}

Thus, the CCC of EX2 is:
\begin{eqnarray*}
    CCC(EX_2) &= w_1*C_1(EX_2)+w_2*(1-C_2(EX_2))+w_3*(1-C_3(EX_2))
    \\ & = 0.4 * 0.98 +0.2*(1-0.66)+0.4*(1-1) = 0.46 
\end{eqnarray*}

\subsection{Condition \& API Call (EX\_3)}
\label{Appenidx-condition}

The following calculations refer to a condition and an API call. For these calculations, we consider the lines of the condition and the print command (the tags of the condition are eliminated from the computations, as they are basically semantic pointers). Thus, the additional code in each function is 2 lines, and the original code in each function consists of 2 lines. 

Thus, we get for $C_1(EX_3)$:
\begin{equation*}
C_1(EX_3) = \sum_{i=1}^{n}\frac{\frac{e^{|l_i|}}{e^{|l_i|}+|s_i|}}{n} = \frac{\frac{e^2}{e^2+2}+\frac{e^2}{e^2+2}}{2}=0.79
\end{equation*}

For $C_2(EX_3)$, as EX3 includes a condition, we get:
\begin{equation*}
C_2(EX_3) =\sum_{i=1}^{n}\frac{C_3(l_i)}{n}=\frac{0.33}{2}+\frac{0.33}{2}=0.33
\end{equation*}

For $C_3(EX_3)$, we get:
\begin{equation*}
C_3(EX_3) =\sum_{i=1}^{n}\frac{C_3(l_i)}{n}=\frac{0.5}{2}+\frac{0.5}{2}=0.5
\end{equation*}

Thus, the CCC of EX3 is:
\begin{eqnarray*}
    CCC(EX_3) &= w_1*C_1(EX_3)+w_2*(1-C_2(EX_3))+w_3*(1-C_3(EX_3)) = \\ 
    & = 0.4 * 0.79 +0.2*(1-0.33)+0.4*(1-0.5) = 0.65 
\end{eqnarray*}

\end{document}